# 0.82 µm, 105 W diode-pumped thulium-doped all-silica-fiber laser

**CHANGSHUN HOU,**[1,*] **ZIWEI ZHAI,**[1] **NILOTPAL CHOUDHURY,**[1] **TOM HARRIS,**[1] **QIUBAI YANG,** [1] **JAYANTA K. SAHU,**[1] **AND JOHAN NILSSON**[1]

[1]*Optoelectronics Research Centre, University of Southampton, Southampton SO17 1BJ, UK*
*\*C.Hou@soton.ac.uk*

**Abstract:** An all-silica-fiber thulium-doped fiber laser emitting at 0.82 µm on the transition from $^3H_4$ to the ground state $^3H_6$ outputs 105 W continuous-wave (CW) power and 555 W quasi-continuous-wave (QCW) instantaneous power with 0.96% duty cycle in 240-µs rectangular pulses. The TDFL comprises a double-clad thulium-doped fiber (TDF) which is designed and fabricated in-house and is incorporated into an all-fiber cavity and cladding-pumped by five pigtailed diode lasers at 0.79 µm. Co-lasing at 1.9 µm counteracts population trapping in $^3F_4$. The slope efficiency relative to absorbed pump power reaches 64% QCW and 77.5% CW. QCW, the beam quality $M^2$ becomes 2.2 (beam parameter product BPP 0.57 mm mrad) and 2.45 (BPP 0.64) in orthogonal directions at ~250 W of instantaneous output power. Additionally, a modified QCW setup is continuously wavelength-tunable from 812 nm to 835 nm. We believe this is the first reported demonstration of high-power laser operation of the $^3H_4 \rightarrow {^3H_6}$ transition in a TDF. Given also the simplicity and other attractions of an all-silica-fiber laser with direct-diode cladding-pumping, we believe our demonstration is valuable for applications ranging from laser machining of aluminum (benefitting from an absorption peak at 0.83 µm) to scientific applications including strontium-based atomic clocks and cesium-based quantum metrology.

## 1. Introduction

An efficient, high-power fiber laser source with high beam quality in the 800 nm spectral region holds strong appeal for applications such as advanced material processing. While ytterbium-doped fiber lasers operating at 1.0 or 1.1 µm have become industry standard for laser machining of metals, their performance is significantly constrained when dealing with materials which are highly reflective in that wavelength range such as aluminum, which plays a crucial role in the aerospace industry and e-mobility. For example, the need for lightweight aircraft structural parts in aerospace is primarily met by aluminum alloys [1], which are much lighter than steel structures [2]. Additionally, aluminum outperforms steel in demanding working conditions, offering superior durability and resilience [3], bringing together properties such as corrosion resistance, strength, and rigidity in a single product [4]. As shown in Fig. 7 of [5], aluminum exhibits reflectance exceeding 95% at 1.06 µm, severely compromising energy deposition at that wavelength. In contrast, shorter wavelengths generally exhibit higher absorption in metals [6]. For instance, reducing the laser wavelength from 1.06 µm to 808 nm approximately triples the aluminum absorption, as also illustrated in Fig. 7 of [5]. This enhanced absorption enables more efficient melting at lower laser powers and opens up for reduced energy consumption. High beam quality can further improve the processing, offering more uniform heating and melting profiles, minimized defects such as cracks, and greater precision in applications like surface treatment and welding [7]. In addition to material processing, high power and high beam quality laser source at 0.8-µm is attractive for advancing quantum technologies. For instance, a high-power 813 nm laser can be utilized for optical lattice trapping in strontium, enabling the creation of stable traps for Sr-based optical lattice clocks (OLCs) with minimal perturbation to clock transitions, thus enhancing precision timekeeping. Higher power levels allow for the collection of more ions, further improving the stability and accuracy of OLCs [8]. Furthermore, a high-power 852 nm laser can be employed for cesium pumping, enabling atomic fountain clocks, interferometry, and focused ion beam (FIB) applications, for which high output power is required to increase the atom number and signal-to-noise ratio (SNR) [9, 10].

As another application, 0.8 µm lasers can be used to pump neodymium-doped fiber lasers emitting near 0.9 µm [11, 12]. Their three-level nature means they benefit from high pump brightness, making them challenging to power-scale when pumping with diode lasers. By contrast, core-pumping with a high-brightness 0.8-µm laser allows for much shorter gain fibers with higher inversion than with cladding-pumping. This reduces the reabsorption and enables operation on shorter wavelengths of the 0.9-µm transition, potentially even shorter than the absorption peak at ~0.89 µm. In addition, single-mode end-pumping of Nd-doped "bulk" traveling-wave optical amplifiers is attractive since it opens up for tight pump-beam confinement and high gain in relatively long gain media, in which a poor beam-quality would necessitate much higher pump powers. This has been demonstrated in Yb:YAG, where a 17-mm-long crystal reached nearly 40 dB of gain at 1030 nm when pumped with 35 W of power at 920 nm from a single-mode neodymium-doped fiber laser [13].

Despite the attraction of these applications, there is a lack of efficient high-brightness 0.8-µm lasers. The power from commercially available single-mode 0.8-µm diode lasers is limited below the watt-level. Though Ti:Al$_2$O$_3$ laser sources can emit at those wavelengths, their high cost and complexity, and their low overall efficiency, limit their practicality. Furthermore, scaling beyond a few tens of watts is challenging.

Cladding-pumped silica fibers doped with Tm$^{3+}$ and operating on the transition $^3H_4 \rightarrow ^3H_6$ (the ground state) have the potential to fill this gap [14]. See Fig. 1 for a partial energy-level diagram and transitions in Tm$^{3+}$. However, the 1100-cm$^{-1}$ maximum phonon energy of silica glass leads to rapid multi-phonon relaxation (MPR) from $^3H_4$ to $^3H_5$. The resulting short fluorescence lifetime of $^3H_4$ (e.g., 10 µs or a few tens of µs), serving as the upper laser level (ULL), hampers population inversion. Consequently, the 0.8-µm transition has received less attention than the well-known 2-µm transition $^3F_4 \rightarrow ^3H_6$, which benefits from a much longer lifetime of its ULL ($^3F_4$) in Tm-doped silica fibers. We also mention that the $^3H_5$ lifetime is short enough, and the energy gap to $^3F_4$ is large enough, to render its population negligible. Theoretical modeling of a silica-based fiber laser at 810 nm has been detailed by Peterka *et al.* [15] and amplification around this wavelength has been demonstrated in core-pumped silica fiber [16]. Furthermore, we have recently reported amplification on the $^3H_4 \rightarrow ^3H_5$ transition at 2.3 µm in Tm$^{3+}$-doped silica fiber [17]. However, before our work, there are no reports of a high-power Tm$^{3+}$-doped fiber laser (TDF laser, TDFL) operating on this transition.

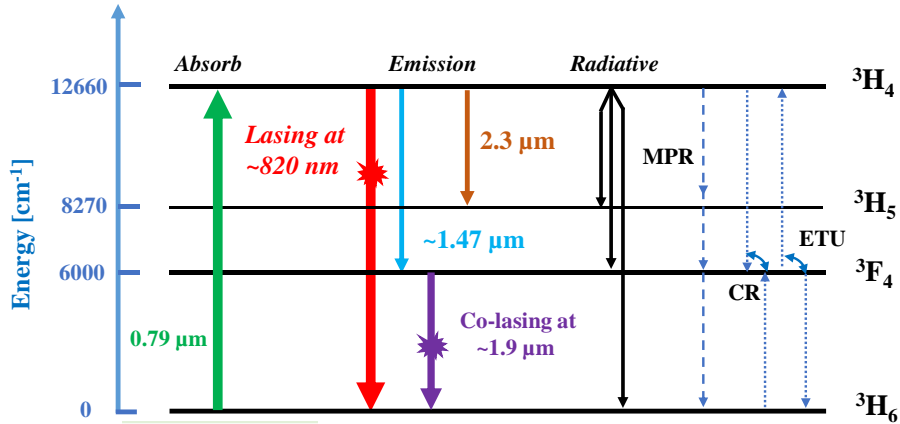

Figure 1. Partial energy level diagram and transitions of Tm$^{3+}$ ions in silica (MPR: multi-phonon relaxation; CR: Cross-relaxation; ETU: Energy transfer upconversion).

To date, research on the 0.8-µm transition has rather focused on fluoride-based fibers [18, 19]. Their lower maximum phonon energy (~500 cm$^{-1}$) leads to a ms-level $^3H_4$ fluorescence lifetime, so up to two orders of magnitude longer than in silica. However, the power-handling of these soft glasses limits power scaling. Currently, the highest 0.8-µm laser power from fluoride-based fibers is at the watt-level [20]. Moreover, soft-glass fibers suffer from fabrication and handling (e.g., splicing) difficulties as well as hygroscopicity and / or low chemical stability. These factors highlight the preference for silica fibers, given their compatibility with existing technologies, durability, and most importantly power scalability, convincingly shown in TDFs at around 2 µm as well as in Yb-doped fibers at 1 – 1.1 µm.

This paper reports our demonstration of 105 W of continuous-wave (CW) and 555 W of instantaneous quasi-CW (QCW) double-ended output power at 0.82 µm from an all-silica-fiber TDFL. The laser was realized using conventional down-conversion with single-step diode-laser-pumping at 0.79 µm and an in-house fabricated TDF. The laser cavity was formed by two perpendicularly cleaved fiber ends. In the QCW regime, we achieved a 0.82-µm slope efficiency of 64% versus absorbed pump power, and 54.2% versus launched power with 240-µs pulses at 40 Hz pulse repetition frequency (PRF) (duty cycle 0.96%). The 0.82-µm beam quality factor $M^2$ was ~2.2 and 2.45 in orthogonal directions. We also demonstrate tuning over the range 812–835 nm in a modified configuration. In the CW regime, the 0.82-µm slope efficiency was 77.5%, and the 0.82-µm optical-to-optical conversion efficiency reached 53.4%, both with respect to absorbed pump power. There was also simultaneous laser emission at ~1.9-µm from the $^3F_4 \rightarrow {}^3H_6$ transition. This prevented the $^3F_4$-level population from building up, which improves the pump absorption [14]. The CW and QCW (instantaneous) output power reached ~25 W and ~54 W, respectively at ~1.9-µm. Compared to our Arxiv posting [21], we here present more than three times the QCW power, CW operation, as well as additional results, details, and discussions.

## 2. Principle of operation

Our approach of single-step cladding-pumping at 0.79 µm with co-lasing at ~1.9 µm involves the levels $^3H_6$, $^3F_4$, and $^3H_4$. The population of other levels is small and will be disregarded. It offers several advantages. Firstly, $^3H_4$ as well as other excited levels are free from excited-state absorption (ESA) at 0.79 µm, improving efficiency and reducing the risk of photodarkening. Secondly, high-power, high-brightness 0.79-µm diode lasers are readily available, providing significant potential for power-scaling. Thirdly, pumping directly to the ULL (in-band) leads to a quantum defect which can be ~4% or even smaller, and thus potentially a high slope efficiency for the 0.8-µm laser. In simulations, the slope efficiency even rivaled that of ytterbium-doped fiber lasers (YDFLs) [22], which except for the co-lasing use similar configurations with in-band pumping. The ~1.9-µm co-lasing is beneficial for the primary $^3H_4 \rightarrow {}^3H_6$ emission at 0.8 µm. It follows on from the nonradiative decay from $^3H_4$, with power dictated by the rate of nonradiative decay from $^3H_4$. It draws on the residual $^3F_4$ energy which is already unavailable for the 0.8-µm lasing and beneficially makes the Tm$^{3+}$-ions available for the pump-absorption step of the 0.8-µm laser cycle. This enhances the pump absorption. Importantly, the 1.9-µm co-lasing does not depopulate $^3H_4$, so does not compete with the 0.8-µm emission and need not impair its slope efficiency or threshold.

## 3. Quasi-continuous-wave operation

We first consider quasi-continuous-wave operation, for which we used pump pulses of 240-µs duration at a PRF of 40 Hz (duty cycle 0.96%).

3.1 *Experimental configuration, quasi-continuous-wave operation*

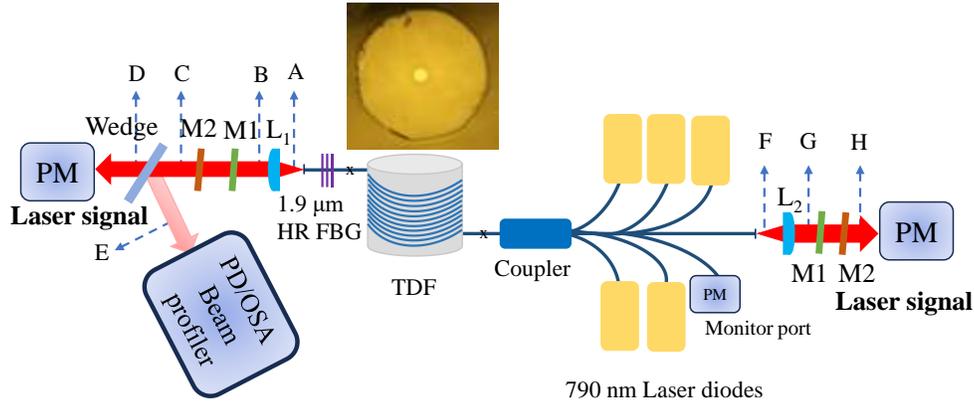

Figure 2. Experimental setup of TDFL. TDF: thulium-doped fiber; FBG: fiber Bragg grating; PM: power meter; PD: photodetector; OSA: optical spectrum analyzer; $L_1$, $L_2$: antireflection-coated collimating lens; M1: short-pass filter, cut-off wavelength: 1600 nm; M2: long-pass filter, cut-on wavelength: 800 nm. A, B, C, D, E, F, G, H: Reference positions for measurements. Inset: microscope image of the TDF (130-µm diameter corner-to-corner).

Fig. 2 illustrates the QCW experimental setup. The TDFL comprises an 18-m-long TDF and five multimode diode lasers emitting at 0.79 µm for pumping. The TDFL emits at 0.82 µm with double-ended output. The length was chosen to yield highest output power, although the length dependence is relatively weak around the optimum. The TDF is coiled on a 10 cm diameter aluminum cylinder, with no special heat dissipation measures required, thanks to the low heat generation at 0.96% duty cycle.

Fig. 2 includes a cross-sectional image of the TDF. This was designed and drawn in-house from a Tm-doped silica preform designed and fabricated in-house. The TDF has a step-index core with diameter ~12 µm and a numerical aperture (NA) of ~0.15, calculated from the preform refractive-index profile. The V-value becomes ~6.9 at 0.82 µm and ~3.0 at 1.9 µm. Surrounding the core is an octagonal pure-silica cladding ("inner cladding") with diameter 130-µm corner-to-corner and 125-µm flat-to-flat. This is coated by a low-index acrylate for an inner-cladding NA of 0.46, nominally. For the shaping, the preform was machined in-house with a $CO_2$-laser prior to fiber draw.

The core propagation loss spectrum was measured with the cutback method. For this, a piece of the TDF was spliced between two pieces of SMF-28, and the transmission of light from a white light source (WLS) was measured with an OSA for two different TDF lengths. The core propagation loss at 0.82 µm is ~0.25 dB/m, comprising both ground-state absorption (GSA) of $Tm^{3+}$-ions and background loss. This is the "cold" loss, with all $Tm^{3+}$-ions in the ground state. At 1.32 µm, the background loss is approximately 0.15 dB/m. This wavelength avoids both $Tm^{3+}$ GSA and $OH^-$-absorption. The $OH^-$ core absorption at 1380 nm was measured to ~0.35 dB/m, from which we estimate the core's $OH^-$-concentration to around 7 ppm.

To assess the pump absorption, the "cold" absorption for light propagating in the inner cladding was measured with a WLS in 15 m of TDF laid out in a loose 10-cm-diameter coil. It reaches ~9 dB at the absorption peak of 787.2 nm (Fig. 3 (a)). Although the cladding edges appear somewhat rounded in the inset of Fig. 2, bending the TDF to mix the pump modes did not enhance the absorption, as it does with circularly symmetric fibers [23]. This suggests that the cladding shaping successfully negated detrimental mode-selective pump depletion.

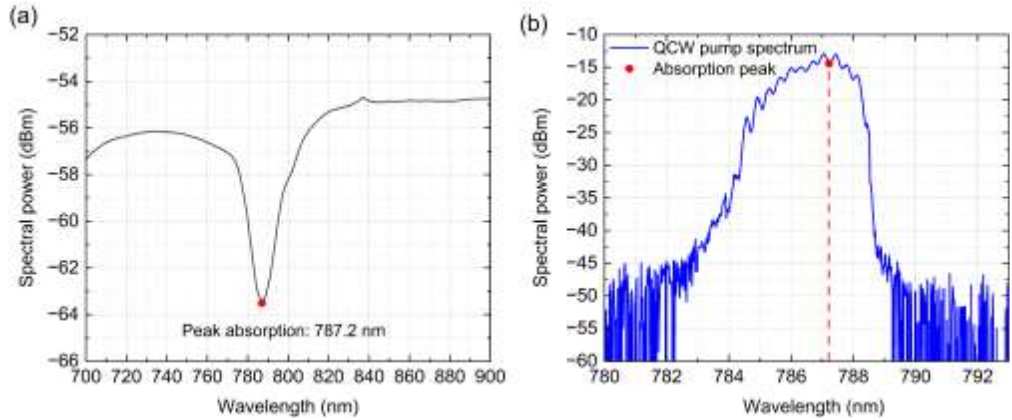

Figure 3. (a) Measured white-light ("cold") cladding transmission spectrum in 15 m of TDF (resolution bandwidth (RBW) 0.1 nm). (b) Combined QCW spectrum of all five 0.79 µm diode lasers at full pump power.

The TDF was cladding-pumped by five 0.79-µm fiber-coupled diode lasers (BWT diodes, K793DL9RN) launched via a (6+1) x 1 pump and signal combiner with six pump ports (Lasfiberio, pump fiber: Coherent MM-S105/125-22A; signal and common-port fiber Coherent SM-GDF-10/130-15M, low index acrylate coating). This was spliced to the TDF. One pump port was left unconnected and instead used to indirectly monitor the backward-propagating 0.82-µm power entering the diode lasers spliced to the other pump ports. This was a precaution, as the diode lasers are only protected internally over the 1900~2100 nm range. Under QCW pumping, the diode lasers were driven in series by diode laser drivers (LDD-1046, 120A/120V, Meerstetter), capable of fast-pulse operation (on & off switching times < 10 µs). The drivers were controlled by 250-µs rectangular pulses at a PRF of 40 Hz from a waveform generator (Tektronix AFG31152, 150 MHz bandwidth, 2 GSa/s sampling rate) via their current control ports. This resulted in rectangular pump pulses of 240-µs duration (duty cycle 0.96%). The pump diodes are rated at 130 W of CW power in a 105-µm core fiber with a NA of 0.22. In QCW operation with 240-µs pulses, they can be driven to produce around ~230 W of instantaneous pump power for a total of ~1.15 kW. The pump transmission of the combiner was measured to ~ 94%. The total maximum instantaneous pump power exiting the combiner into free space was measured to 1.08 kW. Compensating for the Fresnel loss and neglecting splice loss, 1.12 kW of maximum instantaneous pump power was launched into the TDF, limited by available pump power. The emission spectrum of the pump lasers at full power is well matched to the TDF absorption peak (Fig. 3 (a) and (b)). We measured the pump leakage to around –7.87 dB (loss ~0.437 dB/m) at full power.

The 0.82-µm laser cavity was formed by perpendicular fiber cleaves in both ends of the fiber assembly with Fresnel-reflection feedback of 3.5% at 0.82 µm. The laser output power is therefore double-ended. A high-reflecting fiber Bragg grating (FBG) to make the output single-ended was not available. Also for the 1.9-µm laser cavity, the Fresnel reflection (3.2%) from the perpendicularly cleaved fiber end provided the feedback in the pump launch end. In addition, a highly reflecting FBG (TeraXion, fiber: Coherent SM-GDF-10/130-15M, low index acrylate coating) was spliced to the far end of the TDF to close the cavity. Compared to 3.2% Fresnel reflection, this reduces the required $^3F_4 - {}^3H_6$ inversion and advantageously, the number of ions in $^3F_4$.

At both ends of the fiber cavity, antireflection (AR) coated Infrasil aspheric lenses ($L_1$ and $L_2$, NA ~ 0.35, AR range 0.8~2 µm, Laser 2000) collimated the output beams. Although the NA of the lens is smaller than that of the inner cladding, it was able to collect all the light exiting the fiber ends. To separate out the 0.82-µm beams, short-pass filters M1 (1600SP, Edmund Optics) and long-pass filters M2 (FELH800, Thorlabs) removed residual pump and the 1.9-µm co-lasing power. The combination induced a loss of ~0.70 dB at 0.82 µm. In addition, an uncoated wedge was used intermittently to sample the laser output for characterizations including temporal trace, spectrum, and beam quality.

We determined instantaneous QCW powers by measuring average powers with thermal power meters and dividing by the 0.96% duty cycle. Quoted powers and efficiencies for 0.82 µm are based on data measured at positions C and H in Fig. 2 (a), which are then recalculated to positions A and F. The 1.9-µm power was measured in a similar manner by replacing M1 with a long-pass filter (FELH1000, Thorlabs, not shown in Fig. 2 (a)) to directly measure the power afterwards. The value was then recalculated to position F. Thus, the losses in the lenses and filters are compensated for in presented results, by 0.70 dB at 0.82 µm and by and 0.46 dB at 1.9 µm. We also measured the leaked pump, and compensated for its outcoupling losses in the data we report.

Diagnostic equipment of note were a Thorlabs DET10A/M biased silicon PD with 350-MHz bandwidth, a Thorlabs DET10D2 biased InGaAs PD with 14-MHz bandwidth, a Keysight InfiniiVision MSOX3104G oscilloscope with 1-GHz bandwidth, an Ophir FL250A-BB-50-PPS beam tracking power and energy sensor, an Ophir 150-W power meter with a thermopile detector, and a Thorlabs dual-scanning-slit beam profiler (BP209VIS/M) mounted on an $M^2$ measurement system with a motorized translation stage (M2MS). For optical spectra, we used Ando AQ6317B and Yokogawa AQ6376 optical spectrum analyzers.

### 3.2 Quasi-continuous-wave laser results

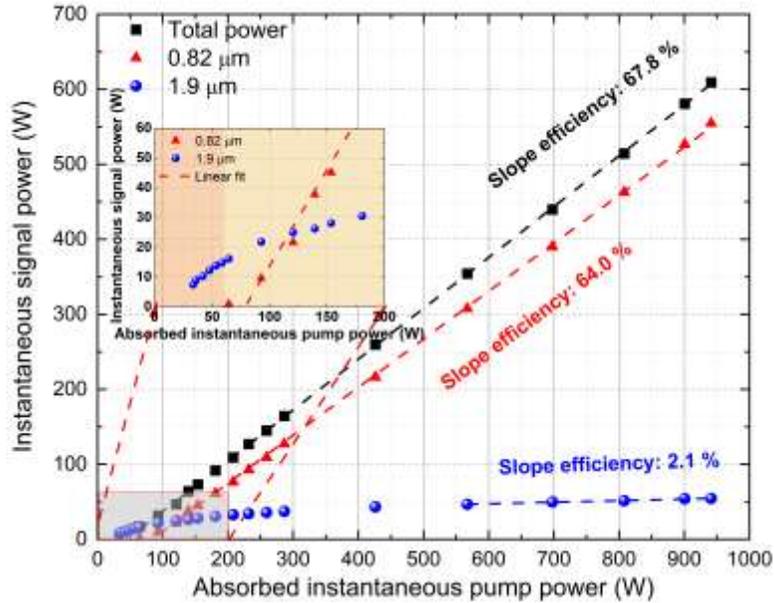

Figure 4. QCW instantaneous output power combined from both ends of the TDFL vs. absorbed instantaneous pump power. Blue dots: power at 1.9 nm; red dots: power at 0.82 µm; black dots: total output power. Inset: enlarged view of the behavior near threshold.

Fig. 4 shows the instantaneous laser output power at 0.82 µm and 1.9-µm as a function of the absorbed instantaneous pump power in the QCW regime. Here and throughout, we take the absorbed pump power to be the difference between launched and leaked pump power (leakage –7.87 dB, i.e., 16.3%, at full power). The 1.9-µm laser threshold is lower than the 0.82-µm threshold. This is understandable, since the lifetime of $^3F_4$ is much longer than that of $^3H_4$. We could not investigate the characteristics below an instantaneous pump power of around 33 W (current of 3.68 A) because the diode drivers were unable to reliably generate lower currents.

The instantaneous absorbed pump threshold power for 0.82-µm laser emission was ~60 W (~75 W launched). The short lifetime of the ULL $^3H_4$ and the high cavity loss contribute to the high threshold, which we believe is among the highest reported for a cladding-pumped fiber laser. Although a high threshold is undesirable, it is still only a small fraction of the available QCW pump power. This underlines the opportunity offered by the high power and brightness of state-of-the-art diode lasers. Thus, at the maximum launched instantaneous pump power of 1.12 kW, ~555 W of instantaneous 0.82-µm output power was obtained, combined from both output ends. The forward and backward powers were nearly the same, which indicates that both ends of the cavity were equivalent and thus that the splices and cleaved fiber facets were good. The 0.82-µm slope efficiency with respect to the absorbed pump power reaches 64%. The output power increases linearly with an increase in the pump power, without any power roll off. It is noteworthy that at full power, ~10 W of instantaneous power, predominantly at 0.82 µm, was measured at the combiner's pump port used for monitoring. This stems from the portion of the generated signal light that leaks into the cladding, for example at splices or as bendloss of higher order modes. Even if it does not contribute to the useful output, it does add to the intrinsic conversion efficiency of the TDF. Assuming that all pump ports contain a similar backward-propagating 0.82-µm power, this adds up to ~60 W which is not included in the output powers or slope efficiencies we report.

Above its threshold, the 1.9-µm output power first increases with slope ~32% with respect to absorbed pump power. The wavelength was centered at around 1901 nm as determined by the FBG. Then, above the 0.82-µm laser threshold, the 1.9-µm slope drops to only 2.1%. Indeed, in an ideal case of a "homogeneous" gain medium with all $Tm^{3+}$-ions being spectroscopically equal and negligible temperature effects, the 0.82-µm and 1.9-µm laser fields clamp the populations of the three populated levels. This in turn clamps the nonradiative decay from $^3H_4$ to $^3F_4$ and thus the stimulated emission from $^3F_4$ to $^3H_6$ in the ideal case. Although 2.1% is small, if this differential loss of this channel were instead funneled into 0.82-µm emission (as it might, with homogeneous gain characteristics), it could add approximately $(1.9 / 0.82) \times 2.1\%$ = 4.86% to the 0.82-µm slope. The instantaneous power at ~1.9-µm reached ~54.4 W at full power, so ~10% of the 0.82-µm power and 18.5% of the total number of photons in the laser output. No degradation in output power was observed during our measurements.

The 0.82-µm output spectrum at different pump power was measured with an optical spectrum analyzer (OSA, Ando, AQ6317B), triggered synchronously with the pulses. See Fig. 5 (a). The output spectrum was centered at 821.6 nm at full power. The spectral width increased linearly with 0.82-µm output power from 3 to 11 nm, as shown in Fig. 5 (b). A large linewidth is expected since the cavity has no spectrally narrow selection for the $^3H_4 \rightarrow {}^3H_6$ emission. The further broadening at higher power may be caused by increased spectral competition as well as by self-phase modulation (SPM), as the nonlinear phase shift is not negligible, according to calculations. Stimulated Raman scattering would occur at ~0.85 µm. This was not observed over the effective noise level of approximately 4 mW in the 0.2-nm resolution bandwidth (determined from Fig. 5 (a)).

Fig. 5 (c) shows temporal traces of the pump and the 0.82-µm and 1.9-µm output. The overall temporal characteristics of the TDFL largely follow those of the diode drivers. The difference in duration and thus duty cycle between the three traces is negligible. Fig. 5 (d) zooms in on the switch-on characteristics. The pump switches on within 10 µs (10%-90%). Relaxation oscillation at 1.9 µm and 0.82 µm occur, respectively, approximately 3 µs and 5 µs after the pump power starts to rise. These die out before the pump power is fully on. Efficient operation remained possible with pulses down to ~100 µs duration.

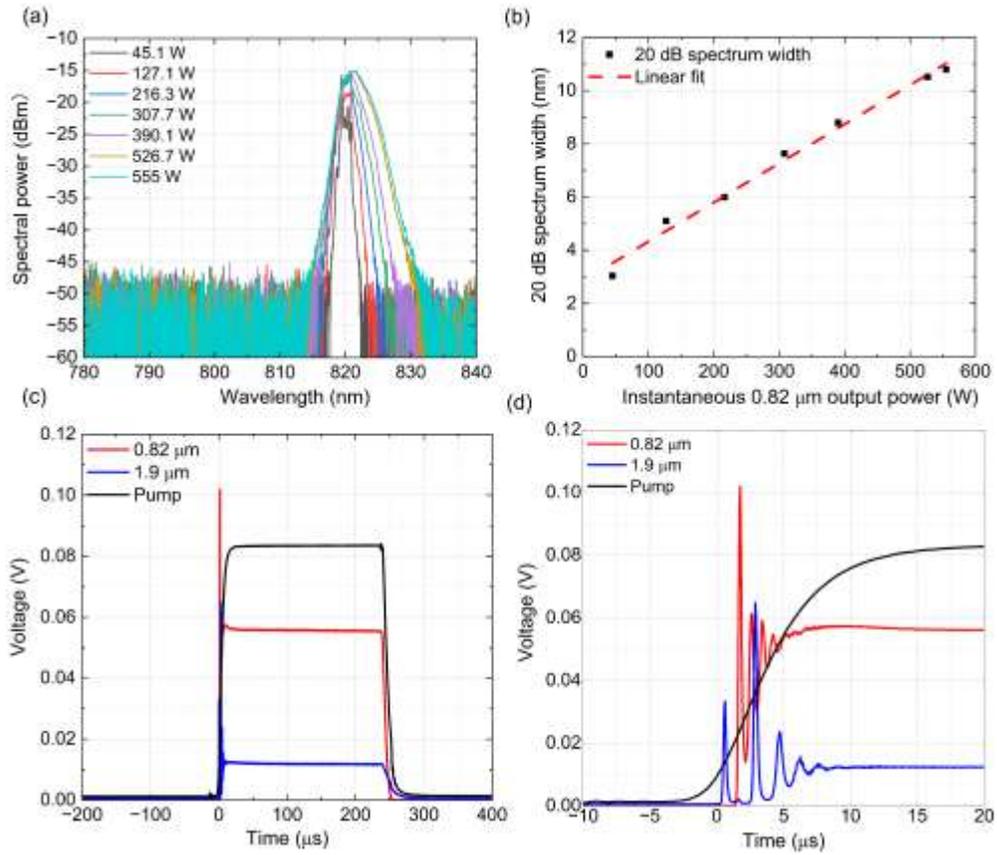

Figure 5. QCW measurements. (a) Optical spectrum at different 0.82-µm output power (RBW: 0.2 nm). (b) 20 dB spectral width vs. 0.82-µm output power. (c) Time-traces of the laser output at 0.82 µm and 1.9 µm and of the 0.79-µm pump at full power. (d) Enlarged view showing relaxation oscillation process.

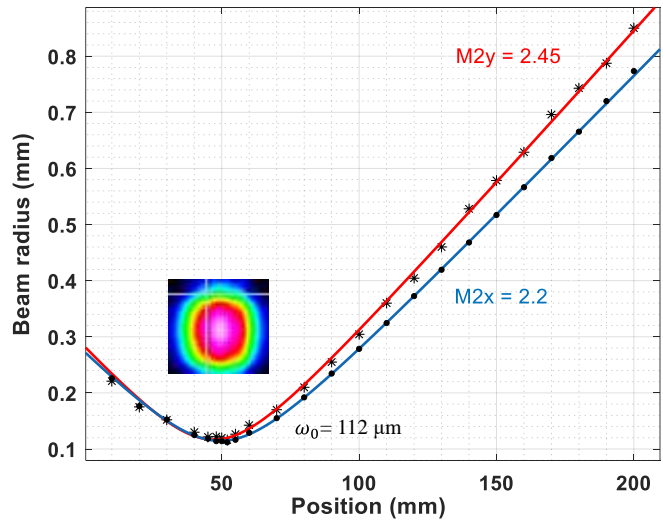

Figure 6. Measured 0.82-µm beam radius in orthogonal directions at different longitudinal positions at ~250 W of instantaneous power. Inset: Synthetic 2D 0.82-µm beam profile as reconstructed by the scanning-slit beam profiler.

The beam quality factor (M²) of the 0.82-µm output was measured at a launched pump power of 640 W with the scanning-slit beam profiler in pulse measurement mode. This was measured in the pump launch end, at Position E in Fig. 2. The results are shown in Fig. 6. The $M^2$-factor was determined to be 2.2 (beam parameter product BPP 0.57 mm mrad) and 2.45 (0.64 mm mrad) in orthogonal directions, based on a hyperbolic fit to the measured beam width at the $1/e^2$ intensity level. This is consistent with the passive output fiber's V-number of 5.75 at 0.82 µm.

Additionally, we investigated the tuning characteristics of the TDFL in the QCW regime. For this, the far output end of a 19-m-long piece of the TDF (near Position A in Fig. 2) was angle-cleaved to suppress feedback. Instead, a Littrow-configured diffraction grating located at Position B with approximately 70% reflectivity at 0.82 µm provided wavelength-selective feedback, tunable by adjusting the angle of incidence. In a non-optimized experiment, the laser then generated between 129 W and 152 W of instantaneous single-ended output power with continuous tuning from 812 to 835 nm for a launched instantaneous pump power of 427 W. See Fig. 7. The slope efficiency was ~50.9%. The 23-nm tuning range corresponds to 10.2 THz, which suggests that Gaussian-shape pulses as short as 40 fs can be amplified. Wider and shifted tuning ranges are expected with tailored TDF parameters and configurations [24].

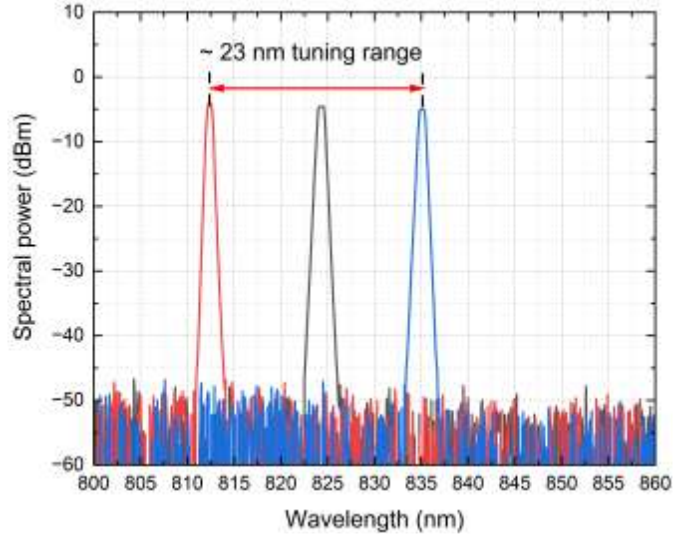

Figure 7. Measured output spectra of the wavelength tunable QCW TDFL at over 100 W of instantaneous output power in the 0.8-µm band (RBW 0.5 nm).

## 4. Continuous-wave operation

4.1 *Experimental configuration, continuous-wave operation*

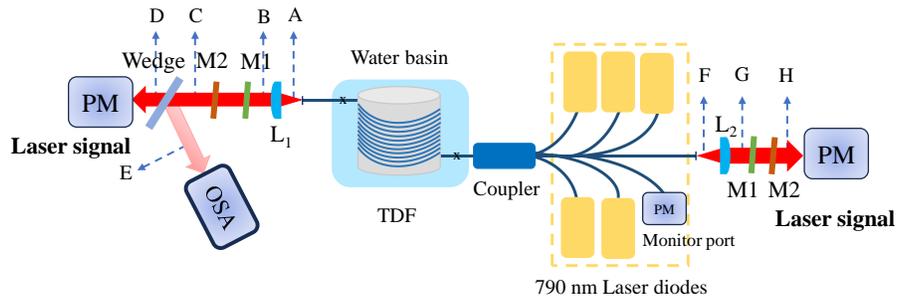

Figure 8. Schematic of CW TDFL with two perpendicularly cleaved ends.

Fig. 8 illustrates the CW experimental setup. The configuration is largely the same as the QCW setup, but with the following differences.

The 0.79-µm diode lasers were driven in series by TDK Lambda Genesys 100-15 power supplies.

Two perpendicular fiber facets were used to form the 1.9 µm cavity as the FBG used QCW failed under high-power CW operation. We believe this was a result of excessive leaked pump power. Thus, we replaced the FBG with a short piece of passive fiber (SM-GDF-10/130-15M, Coherent) spliced to the TDF and with perpendicularly cleaved output end. As a result, the 0.8 µm and 1.9 µm shared the same cavity.

For thermal management, the 0.79-µm diode lasers and pump combiner were mounted on a water-cooled aluminum heatsink. We reiterate that CW, the diode lasers are rated at ~130 W, which is significantly lower than the ~230 W of instantaneous power we used QCW. However, damage at the splice between the pump combiner and the TDF limited the CW pump launch into the TDF to ~291 W of total power (neglecting the splice loss). The power of individual diode lasers was then ~62 W on average before the combiner.

The active fiber, reduced to ~17 m in length due to loss from damage and splice repairs, was wound onto a 13-cm-diameter metal cylinder and fully submerged in water. Also, the splices to the TDF were submerged in order to prevent damage from occurring already at low power. The water temperature was between 15°C and 20°C. These were our primary heatsinking measures. We did not use any heat-conductive foil or paste, or machined grooves to improve thermal contact between the fiber and the cylinder.

Fig. 9 shows individual diode laser spectra at ~130 W of individual power, together with a pump absorption spectrum measured in the same way as in Fig. 3. The spread of the central wavelengths is ~2 nm and the linewidth of individual diode lasers is also ~2 nm. Overall, the spectral match is good, with the diode lasers being marginally too long in wavelength relative to the absorption peak. However, this is at 130 W, whereas the highest used individual diode-laser power was ~62 W. This reduction in power shifts the spectrum to ~3-nm shorter wavelengths. The reduction in absorption due to the spectral mismatch is less than 1 dB from the peak value.

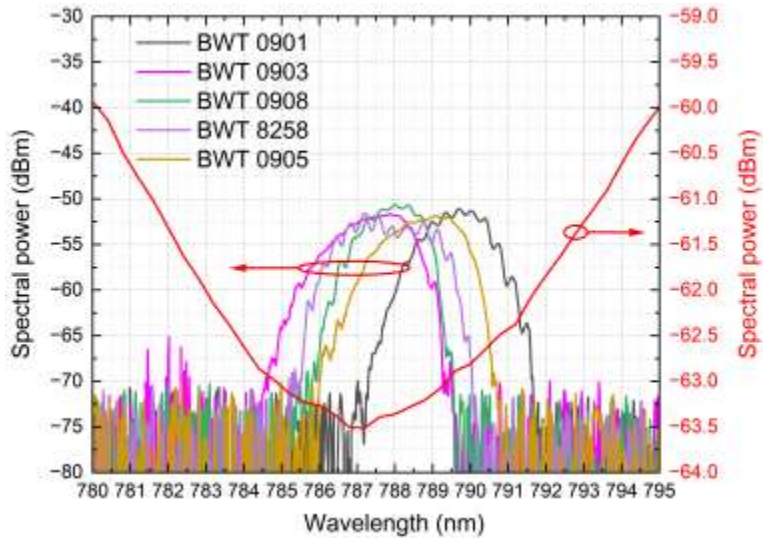

Figure 9. Left axis: wavelength of individual diode lasers. Right axis: Cladding absorption spectrum of the TDF (RBW 0.1 nm).

*4.2 Continuous-wave laser results*

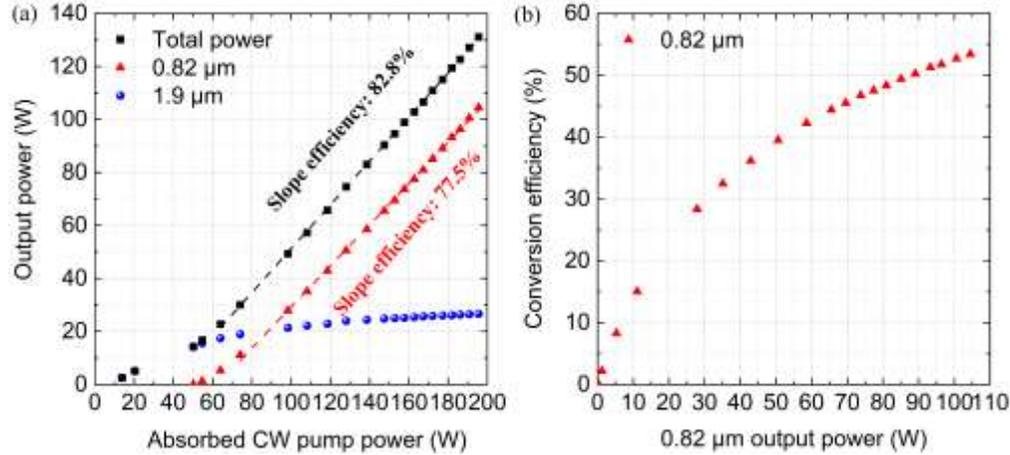

Figure 10. (a) CW output power combined from both ends of the TDFL vs. absorbed pump power. Blue dots: power at 1.9 nm; red dots: power at 0.82 µm; black dots: total output power. (b) Conversion efficiency of 0.82 µm with respect to absorbed pump power vs. 0.82 µm laser output power.

Fig. 10 (a) shows how the measured laser output power depends on the pump power. The pump threshold power for 0.82-µm laser emission with our two 3.5%-reflecting facets was ~50 W (absorbed) and ~67.4 W (launched). The threshold for the 1.9-µm laser emission (now double-ended) is ~10 W. The 0.82-µm output power reached 105 W at the maximum absorbed pump power of 196 W (291 W launched). This was limited by failure at the splice between the pump combiner and the TDF. The slope efficiency of the 0.82 µm emission with respect to absorbed pump power was 77.5%, without any sign of roll-off. This is a notable improvement over the 64% obtained QCW. If we include the 1.9 µm emission, then the total slope efficiency becomes around 83%. This may be a record slope efficiency for any $Tm^{3+}$- doped fiber laser cladding-pumped by diode lasers, or even any diode-laser-pumped $Tm^{3+}$-doped laser, at output power above 100 W.

Fig. 10 (b) presents the optical-to-optical conversion efficiency from absorbed pump to 0.82-µm output power as a function of the output power. At output powers below, say, 20 W, the conversion efficiency suffers from the high laser threshold. However, the conversion efficiency improves to over 50% at an output power of 100 W. FBGs at 0.82 µm may reduce the threshold and thus improve the efficiency at lower power.

The 0.82-µm output spectrum at full power is illustrated in Fig. 11, showing a background which is > 50 dB below the spectral peak as measured with 2 nm RBW. Any stimulated Raman scattering would peak at ~0.85 µm or shorter but was not observed, nor was any spectral broadening caused by the Kerr nonlinearity.

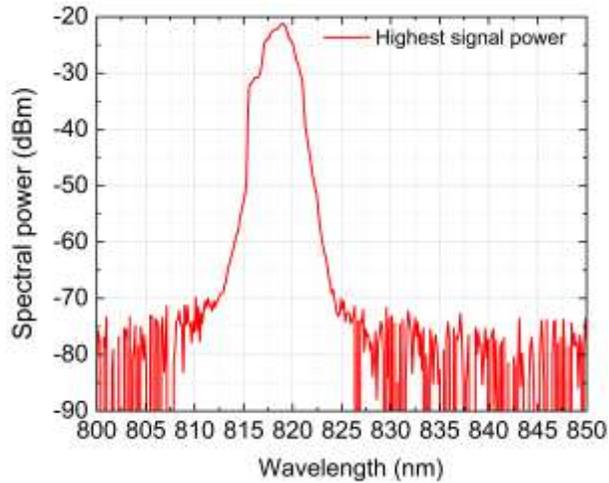

Figure 11. CW output spectrum at full power (RBW 2 nm).

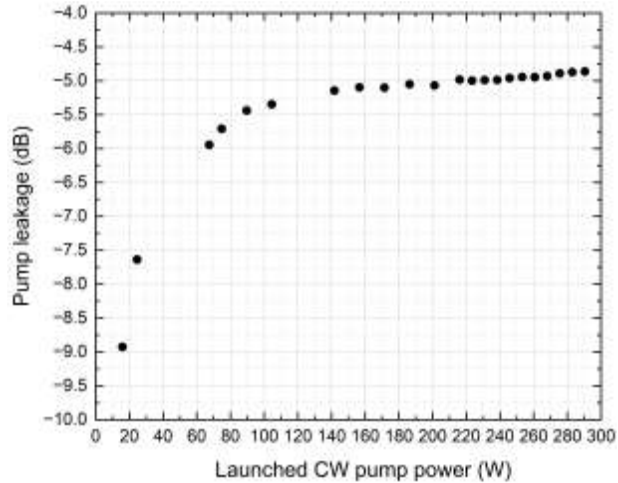

Figure 12. CW pump leakage vs launched pump power.

The 196 W of absorbed pump power at the full launched pump power of 291 W implies a pump leakage of –4.86 dB (32.7%), which is high. Fig. 12 illustrates how the pump leakage varies with launched pump power. At low pump levels, the leakage reaches 9 dB, in good agreement with the WLS transmission spectrum in Fig. 3. As the pump power increases, the leakage rises rapidly, reaching –6 dB at the 0.82-µm threshold of 67.4 W. We attribute this primarily to the reduced population of $Tm^{3+}$ ions in the ground state as the threshold is approached. Beyond threshold, the leakage increases only gradually, by about 1 dB at full power. This slow increase may result from thermal effects in the diode lasers and/or the TDF, which would be consistent with the lower leakage observed in the QCW case. Note that the TDF length was optimized for QCW operation. A longer TDF and improved heatsinking may reduce the pump leakage and allow for higher CW output power.

We operated the system at maximum power for no more than ~10 minutes. Except for catastrophic failures, no noticeable degradation in laser performance (e.g., from photodarkening) was observed. Additional characterization is needed to establish long-term reliability.

## 5. Discussion

Multi-phonon relaxation from $^3H_4$ feeds the $^3F_4$ population, which detrimentally reduces the number of Tm$^{3+}$-ions available for the primary $^3H_6 - {^3H_4}$ transition. To estimate this, we first evaluated the fractional populations of the levels $^3H_6$ ($n_1$), $^3F_4$ ($n_2$), and $^3H_4$ ($n_3$) as averaged over the TDF from the spectroscopic cross-sections and the laser gain thresholds. For this, we used absorption and emission cross-sections $\sigma_a$ and $\sigma_e$ determined on other TDFs. At 0.82 µm, $\sigma_a = 0.244 \times 10^{-25}$ m$^2$ and $\sigma_e = 2.40 \times 10^{-25}$ m$^2$. At 1.9 µm, $\sigma_a = 0.253 \times 10^{-25}$ m$^2$ and $\sigma_e = 3.83 \times 10^{-25}$ m$^2$. In the QCW configuration with an 18-m-long TDF and a cavity roundtrip loss estimated to 30 dB at 0.82 µm and 16 dB at 1.9 µm, we get $n_1 = 0.811$, $n_2 = 0.067$, and $n_3 = 0.123$. The excited ions reduce the "hot" pump absorption by a fraction $1 - n_1 + n_3 (\sigma_e / \sigma_a) = n_2 + n_3 (1 + \sigma_e / \sigma_a) = 0.2986$, approximately, where the cross-section values are those of the pump with ($\sigma_e / \sigma_a$) taken to be 0.8918. Given a "cold" pump absorption of ~9 dB, this suggests a "hot" pump absorption of ~6.3 dB. The difference to the –7.87-dB measured pump leakage may partly be caused by background propagation and splice loss. If $^3F_4$ is empty ($n_2 = 0$) then the "hot" pump absorption becomes around 0.50 dB higher. This suggests that although the finite population of $^3F_4$ reduces the pump absorption, the effect is small and has little impact on the 0.82-µm characteristics in this QCW case.

It is also possible to assess an operating $^3H_4 \rightarrow {^3F_4}$ relaxation rate (dominated by MPR) from the $^3H_4$ population ($n_3 = 0.123$ QCW) and the 1.9-µm power. For example, 20 W of 1.9-µm output power (near the 0.82-µm threshold and generated at a launched pump power of 100 W) corresponds to $1.91 \times 10^{20}$ photons per second. Furthermore, with $n_3 = 0.123$, there are ~$4.99 \times 10^{15}$ Tm$^{3+}$ ions in $^3H_4$. From this, the operating $^3H_4 \rightarrow {^3F_4}$ relaxation rate becomes (26.1 µs)$^{-1}$. At 50 W of 1.9-µm power (launched pump power ~844 W), the operating relaxation rate becomes (10.4 µs)$^{-1}$. This assumes that the $^3H_4$ population is clamped to $n_3 = 0.123$, which it need not be in the presence of inhomogeneities or temperature effects, although the latter are small in the QCW regime.

The (26.1 µs)$^{-1}$ operating relaxation rate is consistent with the initial decay of the 0.8-µm fluorescence following excitation of the $^3H_4$-level by ~100 W, 0.79-µm pump pulses launched into the inner cladding. We measured the fluorescence in the backward direction from a piece of the TDF short enough to avoid reabsorption and stimulated-emission effects. The decay was non-exponential and depended on the pump pulse duration. A pump duration of 20 µs resulted in an initial decay rate of (20.0 µs)$^{-1}$ (1.31 µs after the pump was switched off), a fall time to $1/e$ of 21.5 µs, and a decay rate of (57.2 µs)$^{-1}$ in the long tail. A pump duration of 5 µs resulted in an initial decay rate of (18.1 µs)$^{-1}$ (1.16 µs after the pump was switched off), a fall time to $1/e$ of 19.5 µs, and a decay rate of (52.4 µs)$^{-1}$ in the long tail.

The higher calculated MPR rate of (10.4 µs)$^{-1}$ with 844 W of instantaneous launched pump power may be a result of an increased participation of short-lived ions in the laser cycle at higher power. We also note that our results suggest that the slow decay in the long tail is less relevant. See [25] and references therein for further discussions of the $^3H_4$ lifetime in silica fiber.

The short $^3H_4$ lifetime is a disadvantage. Nevertheless, QCW, the exceptional launched instantaneous pump power of 1.12 kW (~88.4 mW/µm$^2$ intensity) allows us to operate well above the ~75-W threshold (~5.1-mW/µm$^2$ launched intensity) of our configuration, even with single-ended pumping and one pump port unused. However, CW, the threshold is a significant fraction of the failure-limited pump power. Higher power may be possible with improved thermal resilience (e.g., better coating, stripping, or heatsinking) or reduced thermal load per unit length (e.g., lower Tm$^{3+}$-concentration). The diode lasers themselves would allow for over 800 W of single-ended pump power if we use all six pump ports, and ~1600 W with double-ended pumping. However, such high pump power (which would make the threshold negligible) has not been demonstrated with a ~125-µm-diameter pump waveguide and may rather require a diameter of 250 µm or 400 µm. The threshold will be higher, but not overly so in an optimized laser. Note that 400-µm Yb-doped fibers allow for several kW of pump power in commercial high-power lasers.

Even when the pump power is not limited, the complex spectroscopy of $Tm^{3+}$ contains many potential channels for loss (including nonlinear loss), which may also hamper power-scaling. Nevertheless, the lack of roll-off QCW as well as CW does indicate higher power is possible. The potential seems significant, and if the low 4% quantum defect is fully reflected by a low thermal load, then multi-kW CW operation may well be possible.

Operation at lower power (say, 10 W) is also of interest, but in this regime, the high threshold hampers the efficiency. See Fig. 10 (b). A smaller pump waveguide (e.g., 80 µm), a smaller core, and longer $^3H_4$-lifetime can help to reduce the threshold. It is also possible to reduce the outcoupling and thus the cavity loss, although this typically also reduces the slope efficiency. We also note that QCW operation enables high efficiency at low average power.

## 6.  Conclusion

We have demonstrated ~105 W of output power CW and ~555 W of instantaneous output power QCW at 0.82 µm from an all-silica-fiber thulium-doped fiber laser. The $Tm^{3+}$-doped fiber was designed and fabricated in-house and was cladding-pumped by pigtailed diode lasers at 0.79 µm. Population trapping in $^3F_4$ was counteracted by co-lasing at 1.9 µm. Under QCW pumping with pulse duration of ~240 us and repetition rate of 40 Hz, the 0.82-µm slope efficiency reached 64% versus absorbed pump power, and 54.2% versus launched power. The measured beam quality $M^2$ at ~250 W instantaneous output power was 2.2 (BPP 0.57 mm mrad) and 2.45 (BPP 0.64 mm mrad) in orthogonal directions. Moreover, a modified configuration with a diffraction grating was continuously tunable from 812 nm to 835 nm.

In CW operation, the 0.82-µm slope efficiency was 77.5% with respect to absorbed pump power. The achieved output power of 105 W at 0.82 µm was limited by failure at a splice. The high multi-phonon relaxation rate of the upper laser level $^3H_4$ leads to a high threshold of ~50 W of absorbed pump power. However, at high power, the impact of this on the conversion efficiency is small, thanks to the use of high-power, high-brightness pump diode lasers. Thus, the optical-to-optical conversion efficiency reached 53.4% with respect to absorbed pump power. The single-pass 0.82-µm gain was as high as 15 dB in our configuration, thus pointing to the possibility of a 0.82 µm fiber amplifier offering simultaneous high gain, high-power, and high efficiency.

The low quantum defect (e.g., 4%) is a primary attraction of a 0.82-µm TDFL. If this is fully reflected by a low thermal load, then CW multi-kW operation may well be possible. To the best of our knowledge, this is the first experimental demonstration of a high-power silica-based fiber laser source operating in the 800-nm range. We believe these results enable a broad range of applications through a robust and reliable diode-pumped all-fiber laser source.

**Funding.** Air Force Office of Scientific Research (FA9550-17-1-0007), EPSRC EP/T012595/1 (LITECS)

**Acknowledgment:** Changshun Hou thanks the China Scholarship Council for the financial support (Grant No. 202206160003). Jiaying Wu processed the fluorescence decay data.

**Disclosures.** The author declares no conflicts of interest.

**Data availability.** All data supporting this study are openly available from the University of Southampton repository at https://doi.org/10.5258/SOTON/D3392.